\documentclass[12pt,preprint]{aastex}

\shorttitle{Properties of WOH G64}
\shortauthors{Levesque et al.}

\begin{document}

\title{The Physical Properties of the Red Supergiant WOH G64: The Largest Star Known?$^1$}
\author{Emily M. Levesque$^{2}$}
\affil{Institute for Astronomy, University of Hawaii, 2680 Woodlawn Dr., Honolulu, HI 96822}
\email{emsque@ifa.hawaii.edu}

\author{Philip Massey}
\affil{Lowell Observatory, 1400 West Mars Hill Road, Flagstaff, AZ 86001}
\email{phil.massey@lowell.edu}

\author{Bertrand Plez}
\affil{GRAAL, Universit\'{e} Montpellier, CNRS, 34095 Montpellier, France}
\email{bertrand.plez@graal.univ-montp2.fr}

\author{Knut A. G. Olsen}
\affil{National Optical Astronomy Observatory, 950 North Cherry Avenue, Tucson, AZ 85748}
\email{kolsen@noao.edu}

\footnotetext[1]{This paper is based on data gathered with the 6.5 m Magellan telescopes located at Las Campanas, Chile.}
\footnotetext[2]{Predoctoral Fellow, Smithsonian Astrophysical Observatory, 60 Garden St., Cambridge, MA 02138}

\begin{abstract}
WOH G64 is an  unusual red supergiant (RSG) in the Large Magellanic Cloud (LMC), with a number of properties that set it apart from the rest of the LMC RSG population, including a thick circumstellar dust torus, an unusually late spectral type, maser activity, and nebular emission lines. Its reported physical properties are also extreme, including the largest radius for any star known and an effective temperature that is much cooler than other RSGs in the LMC, both of which are at variance with stellar evolutionary theory. We fit moderate-resolution optical spectrophotometry of WOH G64 with the MARCS stellar atmosphere models, determining an effective temperature of 3400 $\pm$ 25 K. We obtain a similar result from the star's broadband $V-K$ colors. With this effective temperature, and taking into account the flux contribution from the aysmmetric circumstellar dust envelope, we calculate log($L/L_{\odot}$) =  5.45 $\pm$ 0.05 for WOH G64, quite similar to the luminosity reported by Ohnaka and collaborators based on their radiative transfer modeling of the star's dust torus. We determine a radius of $R/R_{\odot}$ = 1540, bringing the size of WOH G64 and its position on the H-R diagram into agreement with the largest known Galactic RSGs, although it is still extreme for the LMC. In addition, we use the \ion{Ca}{2} triplet absorption feature to determine a radial velocity of 294 $\pm$ 2 km s$^{-1}$ for the star; this is the same radial velocity as the rotating gas in the LMC's disk, which confirms its membership in the LMC and precludes it from being an unusual Galactic halo giant. Finally, we describe the star's unusual nebula emission spectrum; the gas is nitrogen-rich and shock-heated, and displays a radial velocity that is significantly more positive than the star itself by 50 km s$^{-1}$.
\end{abstract}


\section{Introduction}
\label{sec:intro}
Red supergiants (RSGs) are an evolved He-burning phase in the life of a moderately massive star (10-25 M$_{\odot}$). Their location in the H-R diagram has, in the past, been at odds with the predictions of stellar evolutionary theory, occupying a position that was too cold and too luminous to agree with the position of the evolutionary tracks. Levesque et al.\ (2005, 2006; hereafter Papers I and II) and Massey et al.\ (2009) have used the newest generation of the MARCS stellar atmosphere models (Gustafsson et al.\ 1975, 2003, 2008; Plez et al.\ 1992; Plez 2003) to fit model spectral energy distributions to moderate-resolution spectrophotometry of red supergiants in the Milky Way, Magellanic Clouds, and M31. The new physical parameters derived from this fitting shifted the properties of RSGs to higher effective temperatures and lower luminosities, bringing them into generally excellent agreement with the evolutionary tracks.

However, the RSGs includes several oddball members that require a more detailed analysis. One classic example is the case of VY CMa, a well-studied dust-enshrouded RSG in the Milky Way. Massey et al.\ (2006) fit a spectrum of VY CMa with the MARCS models and derived an effective temperature (T$_{\rm eff}$) that was much warmer than previous estimates (Le Sidaner \& Le Bertre 1996) and brought this previously troublesome star into much better agreement with the evolutionary tracks; however, determination of VY CMa's stellar luminosity remains difficult due to the challenges posed by the large structured dust-reflection nebula that surrounds the star (Monnier et al.\ 1999; Smith et al.\ 2001; Smith 2004, Humphreys et al.\ 2005; Massey et al. 2008) and its uncertain distance (Choi et al.\ 2008). Another example is the unusual Magellanic Cloud RSGs with variable effective temperatures and dust production as described in Levesque et al.\ (2007) and Massey et al.\ (2007), and henceforth referred
to as Levesque-Massey variables. These stars show considerable changes  in their effective temperatures and $V$ magnitudes and a common set of variations - when they are at their warmest they are also brighter, dustier, and more luminous. They are recognizable in their cooler states by occupying the ``forbidden" region on the H-R diagram to the right of the Hayashi track, where stars are no longer expected to be in hydrostatic equilibrium (Hayashi 
\& Hoshi 1961). They are currently believed to be in an unstable (and short-lived) evolutionary phase not previously observed in RSGs.

WOH G64 ($\alpha_{\rm 2000}=$04:55:10.49, $\delta_{\rm 2000}$=-68:20:29.8, from Buchanan et al.\ 2006) is another unusual red supergiant in the Large Magellanic Cloud (LMC), originally discovered by Westerlund et al. (1981), with a number of properties that set it apart from the rest of the LMC RSG population. It is an IRAS source (IRAS 04553-6825), and is surrounded by an optically thick dust torus (Elias et al.\ 1986, Roche et al.\ 1993, Ohnaka et al.\ 2008). Studies of the optical spectra by Elias et al.\ (1986) and van Loon et al.\ (2005) show a spectrum dominated by very strong TiO bands, which has led to assignments of spectral types as late as M7-8, by far the latest spectral type assigned to an LMC RSG (cf., Paper II), with corresponding extreme physical properties, which would make WOH G64 the star with the largest known radius, and the RSG with the largest known luminosity and mass (van Loon et al.\ 2005). It is also a known source of OH, SiO, and H$_2$O masers (Wood et al.\ 1986, van Loon et al.\ 1996, 1998, 2001), with two different maser components, suggesting a substantial mass outflow and two different expanding dust shells (Marshall et al.\ 2004). Finally, Elias et al.\ (1986) report detections of H$\alpha$, [OI], and [NII] emission in the optical spectrum. WOH G64 is neither isolated nor in a well populated region. The star is not in any of the Luck-Hodge OB associations (Lucke \& Hodge 1970).  Inspection of the UBVR  CCD images of the LMC described by Massey (2002) reveals that although there are bright neighboring blue stars in the vicinity (such as [M2002] 22861, located 70" away, or the star BI 23, classified as B0 II by Conti et al.\ 1986, located 110" away), the region is not rich in bright, blue stars.

We have investigated this unique RSG in more detail, applying a similar analysis used in Papers I and II to examine the physical properties that can be derived from the optical spectrum. We obtain moderate-resolution spectrophotometry of WOH G64 using IMACS on the Baade 6.5m Magellan telescope at Las Campanas Observatory (\S~\ref{sec:obs}), confirm this star's membership in the LMC (\S~\ref{sec:mem}), and compare the parameters that we determine from our model fitting to those presented in past work and predictions of evolutionary theory (\S~\ref{sec:prop}). Finally, we discuss the intriguing issue of emission line features in WOH G64 (\S~\ref{sec:elines}) and summarize our findings on this unusual star (\S~\ref{sec:disc}).

\section{Observations}
\label{sec:obs}
A flux-calibrated spectrum of WOH G64 was obtained on UT 2008 December 10 using the IMACS spectrometer
on the Baade 6.5-m Magellan telescope during an observing run primarily devoted to an unrelated
project.  A 300 line mm$^{-1}$ grating was used in first order for spectral
coverage from 3500\AA\ to 9500 \AA. A 0.9"-wide long slit was used, resulting in a spectral resolution of 4.3\AA.
The observation consisted of 3x600 sec exposures, and were obtained under photometric conditions.  The seeing,
measured in the visible off the guide camera (which is off-axis) registered 0.5-0.6" during the exposure. We measure 0.7" full width half maximum intensity on our spectra. The
airmass was approximately 1.5 at the time.  The slit was oriented to the parallactic angle; we note that IMACS
also includes an atmospheric dispersion compensator, so our relative fluxes from the red to the blue should be
good despite the modestly large airmass.

Flat-fielding was performed by means of a series of flats obtained on a projector screen; wavelength calibration (He-Ne-Ar
lamps) were obtained immediately after the three exposures.  A large number of bias frames (60) were averaged
and used to remove residual two dimensional structure after overscan correction and trimming.

In addition to the observation of WOH G64, observations of three LMC RSGs with known radial velocities
from Massey \& Olsen (2003) were obtained immediately prior to the WOH G64 observation in order to
check on LMC membership for WOH G64 (see \S~\ref{sec:mem}).  Those exposures were 60 sec long, and obtained at the parallactic angle. Each observation
was also followed by a HeNeAr comparison arc.

Throughout the course of this and several other nights we also observed spectrophotometric standard stars from
the list of Hamuy et al.\  (1994). The spectrograph slit was oriented to the parallactic angle for these observations as well. Some of those data were obtained under non-photometric conditions, but all of
the standards data were combined, with a grey shift calculated such that the average through-put agreed with the observation with the highest through-put. The resulting residuals from a high-order
fit were a few hundrenths of a magnitude, as expected.  The sensitivity curve we derived was applied to the WOH G64
observation.

All data were processed using IRAF\footnotemark[1]. \footnotetext[1]{IRAF is distributed by NOAO, which is operated by AURA, Inc., under cooperative agreement with the NSF.} The spectra were extracted using an optimal extraction algorithm, with deviant pixels identified and rejected based upon the assumption of a smoothly varying spatial profile.

 \section{Membership in the LMC}
 \label{sec:mem}Before investigating the physical properties of WOH G64, we first wished to confirm that it was indeed a RSG member of the Magellanic Clouds, rather than an unusual foreground giant. Elias et al.\ (1986) measure a radial velocity of 315 km s$^{-1}$ from several nebular emission lines, but state that the velocity zero point is uncertain by 50 km s$^{-1}$, and cite this as evidence of membership in the LMC. However, this really only demonstrates that the star is not a foreground dwarf; a halo giant will have a similar velocity to the LMC's systematic velocity, as it is due mainly to the reflex motion of the sun. This value can also be compared to the $\sim$ 270 - 278 km s$^{-1}$ velocity of the maser emission reported by Marshall et al.\ (2004), which is more in line with our expectations of the LMC's rotation (Olsen \& Massey 2007). Finally, van Loon et al. (1998) use the Ca II triplet to measure a heliocentric velocity of $\sim$ 300 km s$^{-1}$.

\begin{figure}
\epsscale{1}
\plotone{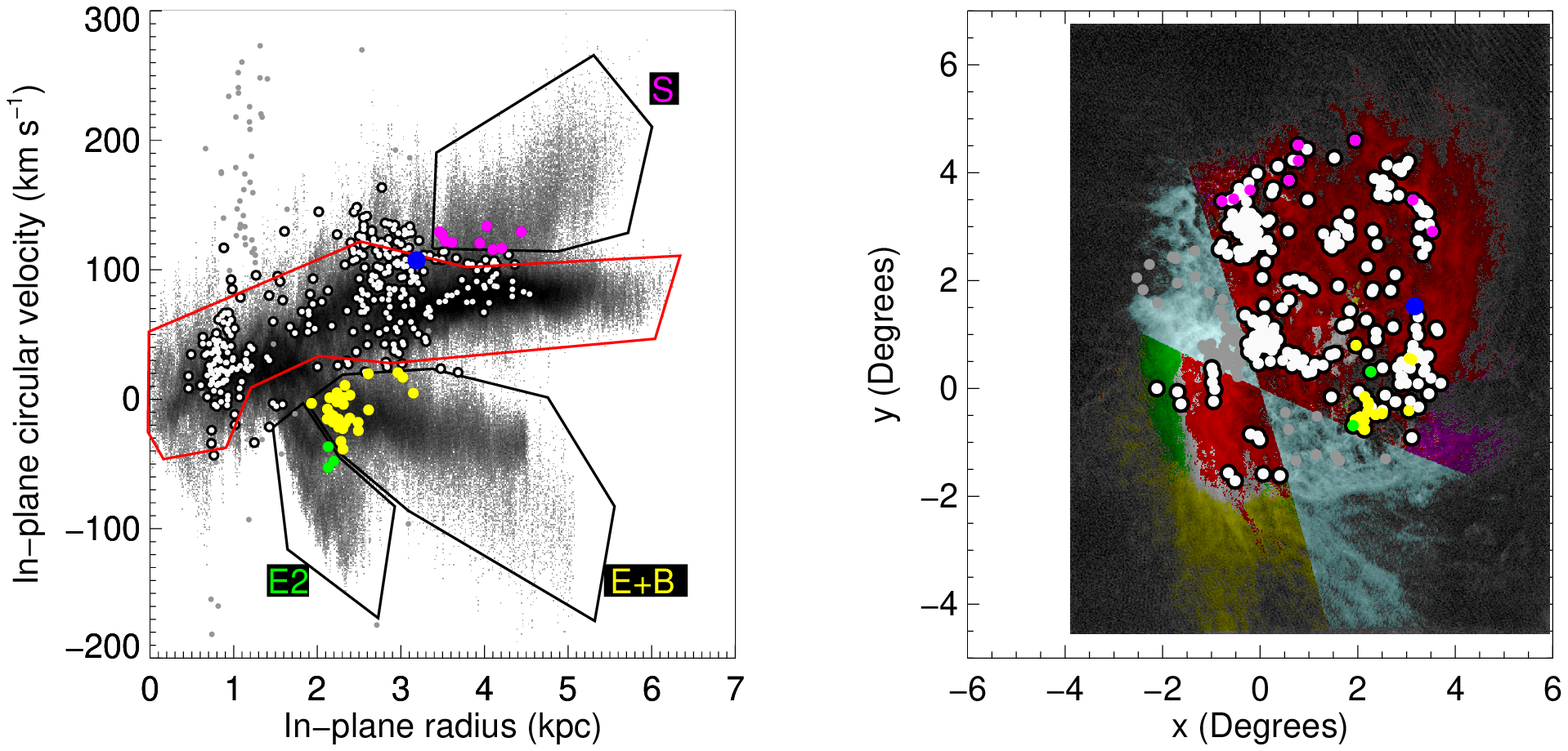}
\caption{\label{fig:kinematics} The velocity of WOH G64 compared to the kinematics of the LMC's HI gas and other LMC RSGs. Left: Measured HI velocities (grayscale; Kim et al.\ 2003) and LMC RSGs (points; Massey \& Olsen 2003) have been converted to in-plane circular velocities, as described in Olsen \& Massey (2007), and are plotted vs. in-plane radius. The polygons labeled ``S'', ``E+B'', and ``E2'' mark the signatures of HI tidal streamers identified by Staveley-Smith et al.\ (2003) and Olsen \& Massey (2007); different colors (green, yellow, and magenta) identify RSGs that fall within these regions. The red polygon outlines the LMC's flat internal rotation curve. The large blue dot marks WOH G64, which clearly has a velocity typical of RSGs at its in-plane radius. Right: The H I gas contained within the regions drawn at bottom left are plotted with different colors as follows: red for the main rotation curve, magenta for the arm S region, yellow for the combined arm E and B regions, and green for region E2. The blue dot again marks the location of WOH G64, while the remaining points are other LMC RSGs. WOH G64 is coincident with the gas that is rotating in the LMC's disk. }
\end{figure}

We can use our spectra to determine if the star's radial velocity is consistent with the kinematics of other RSGs in the LMC. We measured the radial velocity of WOH G64 by cross correlating our spectra against those of three LMC RSGs with accurate velocities measured by Massey \& Olsen (2003), [M2002] LMC 022204, 024014, and 024987. For this, we used the \texttt{fxcor} task in the IRAF \texttt{rv} package, restricting the cross-correlation region to 8450-8700A, which contains the very strong \ion{Ca}{2} triplet lines at $\lambda \lambda 8498, 8542, 8662$ (Figure 1). The spectra were first normalized to unity, and then a value of 1 subtracted so that only spectral features add signal to the cross-correlation. We used the three individual WOH G64 spectra in order to determine the internal consistency of our measurements; in addition we cross-correlated each of the radial velocity ``standards" against each other. These tests revealed that the maximum difference seen in cross-correlating any two calibrating stars against
themselves was 10 km s$^{-1}$, giving us assurance that the velocity we determine with
the combination of the three would be a few km/sec. We expect to be able to centroid each line to about 10\% of the spectral resolution element or better, and so this is consistent since the 4.3\AA\ resolution translates to about 150 km s$^{-1}$. We obtained a radial velocity of 294 $\pm$ 2 km s$^{-1}$, where the error refers to the formal standard deviation of the mean, and reflects
the good agreement in cross-correlating the stars used
for the velocity determinations against themselves. Given this star's location in the LMC, we expect a heliocentric velocity of 263 km s$^{-1}$ due to contributions from the LMC's space motion relative to the sun and its internal rotation, using a kinematic model fit to LMC carbon stars (Olsen \& Massey 2007). Although our measured velocity for WOH G64 is $\sim$30 km s$^{-1}$ higher than that of the carbon stars, it is still entirely consistent with LMC membership, as seen in Figure \ref{fig:kinematics}. The difference in velocity can be attributed to the fact that the LMC's RSGs appear to exhibit $\sim$45 km s$^{-1}$ faster rotation than the carbon stars (whether this is because either the LMC's RSGs or its carbon stars have been kinematically disturbed by external forces is not yet clear; see Olsen \& Massey 2007). In any case, WOH G64's velocity is indeed typical for an LMC RSG at its position.  

\section{Physical Properties}
\label{sec:prop}
\subsection{Previous Work}
\label{sec:past}
Elias et al.\ (1986) obtain a 6200\AA\  - 9200\AA\ spectrum of WOH G64 from the CTIO Blanco 4-meter and RC Spectrograph. Based on this spectrum they assign the star a type of $\sim$ M7.5, and derive an M$_{\rm bol}$ of $-9.7$ (adopting a distance modulus of 18.6) by integrating over IR ground-based and IRAS data and assuming the ($V-K$) color to be $\sim$ 8; we correct this to M$_{\rm bol} = -9.6$ ($L/L_{\odot} = 5.4 \times 10^5$) adopting a distance modulus of 18.5 (van den Bergh 2000). This work is also the first to note the incredible excess IR emission from WOH G64 and postulate that such emission is coming from a substantial circumstellar dust shell.

van Loon et al.\ (2005) examine a low-resolution long-slit optical spectrum obtained with DFOSC at the 1.5m Danish telescope at La Silla in December 1995. They assign a spectral type of M7.5e I, in agreement with Elias et al.\ (1986), but also find some disagreement between different TiO bands, with the 6200\AA\ band suggesting an M7-8 I type while some relative band
strengths suggest an earlier type of M5 I. They assign a T$_{\rm eff}$ of 3008 K based on a giant-class T$_{\rm eff}$ scale published in Fluks et al.\ (1994) and use of the dust radiative transfer model DUSTY (Ivezi\'{c}, Nenkova, \& Elitzur 1999). Finally, they determine M$_{\rm bol} = -9.49$ ($L/L_{\odot} = 4.90 \times 10^5$), adopting a distance to the LMC of 50 kpc.

Most recently, Ohnaka et al.\ (2008) use 2005 and 2007 N-band observations to compute 2D models of the dusty torus surrounding WOH G64. While Ohnaka et al.\ (2008) begin by discussing the parameters derived by van Loon et al.\ (2005) and Elias et al.\ (1986), the temperatures cited within are not derived from the latter papers; instead, Ohnaka et al.\ (2008) correlates the spectral type range of M5-7 with T$_{\rm eff}$ values of 3200 - 3400 K. Adopting the 3200 K T$_{\rm eff}$ in their radiative transfer modeling, and a distance to the LMC of 50 kpc, they determine an M$_{\rm bol}$ of $-8.9$ ($L/L_{\odot} = 2.8 \times 10^5$) based on radiative transfer modeling of the dusty torus around the star. They also compute 2D models with T$_{\rm eff}$ = 3400 K and find that this affects the resulting parameters very little (Ohnaka et al. 2008). The final luminosity determined from this modeling is a factor of 2 lower than Elias et al.\ (1986) and van Loon et al.\ (2005)'s luminosities.

\subsection{Properties from Spectral Fitting}
\label{sec:fit}
Based on the strengths of the TiO bands in our WOH G64 optical spectrum, we assign a spectral type of M5 I by comparison with other RSGs we have classified (Papers I and II; Massey et al.\ 2009). This type confirms WOH G64 as the latest-type RSG in the LMC, although it is considerably earlier than the M7.5 types discussed in Elias et al.\ (1986) and van Loon et al.\ (2005).

The observed SED for WOH G64 was compared to MARCS stellar atmosphere models of metallicity $Z/Z_{\odot} = 0.5$, corresponding to the metallicity of the LMC (Westerlund et al.\ 1997, Massey et al.\ 2004). The MARCS model SEDs available ranged from 3000 to 4500 K in 100 K increments, and were interpolated for intermediate temperatures at 25 K increments. The log $g$ values ranged from $-1$ to $+1$ in increments of 0.5 dex. When fitting the models to the data, they were reddened adopting the Cardelli et al.\ (1989) $R_V = 3.1$ reddening law.

The reddening and T$_{\rm eff}$ were determined by finding the best by-eye fit between the MARCS models and the WOH G64 SED. The reddening was based on the agreement between the model and observed continuum, while T$_{\rm eff}$ was based on the strengths of the TiO bands at $\lambda\lambda$6158, 6658, 7054 (Jaschek \& Jaschek 1990), with the bluer TiO bands at $\lambda\lambda$5167, 5448, 5847 used as secondary confirmations of the fit quality; this ensured that there was minimal degeneracy in determining the best model fit. The $\lambda$8433 and $\lambda$8859 TiO bands are present in the spectrum as well; however, we have found in past work that these features are not generally well-matched by the MARCS models (see Papers I and II). Considering WOH G64's strong TiO band strengths and corresponding late spectral type, our precision for this fit was $\pm$ 25 K. The extinction value $A_V$ is determined to approximately 0.15 mag precision.

\begin{figure}
\epsscale{1}
\plotone{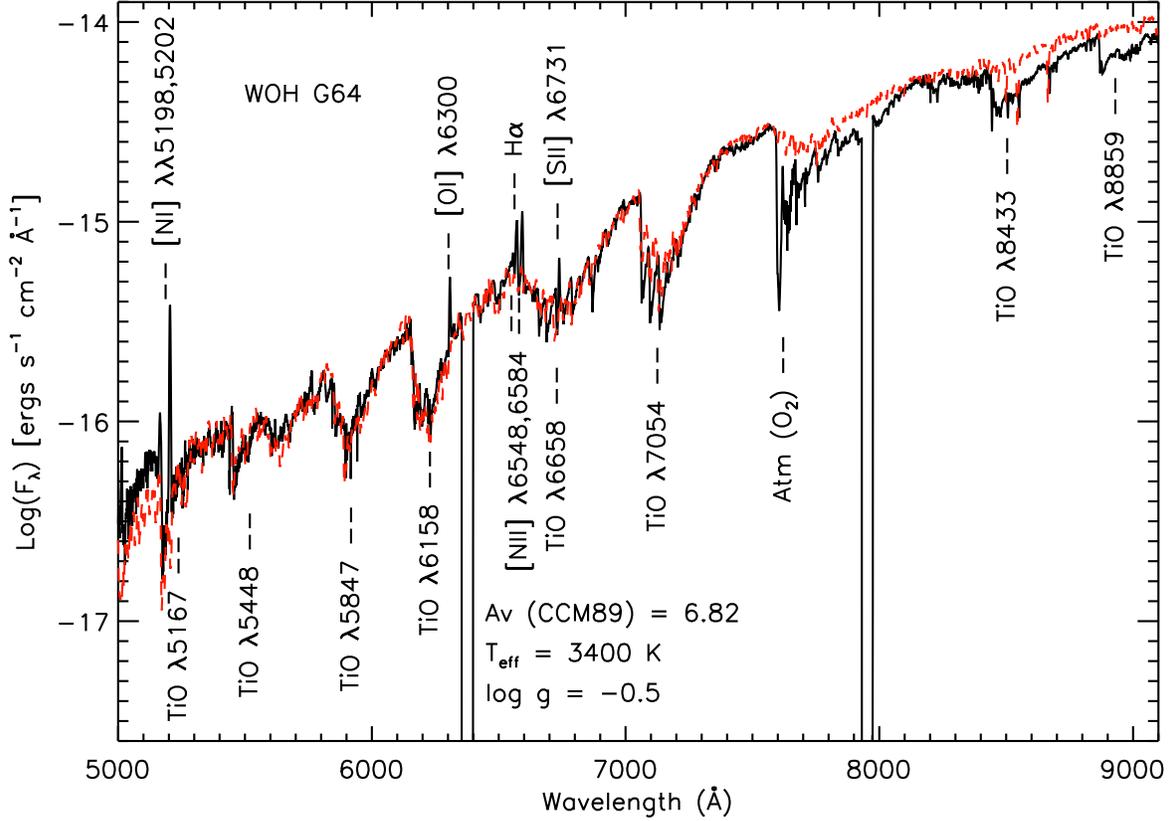}
\caption{\label{fig:spec} Optical spectral energy distribution of WOH G64 (black) overplotted with the best-fit MARCS stellar atmosphere model with T$_{\rm eff}$ = 3400 K (red); there is good agreement with the depths of the TiO bands at 5167\AA, 5448\AA, 5847\AA, 6158\AA, 6658\AA, and 7054\AA, as well as with the overall continuum. Agreement with the 8433\AA\ and 8859\AA\ lines is less satisfactory; however, this is consistent with our past experience that these features are not well-matched with the MARCS models (Papers I and II). Nebular emission lines detected in the spectrum are labeled here and include [NI], [OI], H$\alpha$, [NII], and [SII]. The gaps in the spectrum at $\sim$ 6350\AA\ and 7900\AA\ are the result of gaps in the IMACS CCD mosaic and do not interfere with our model fitting. The strong feature at 7600\AA\ is the telluric A band.}
\end{figure}

\begin{figure}
\epsscale{0.5}
\plotone{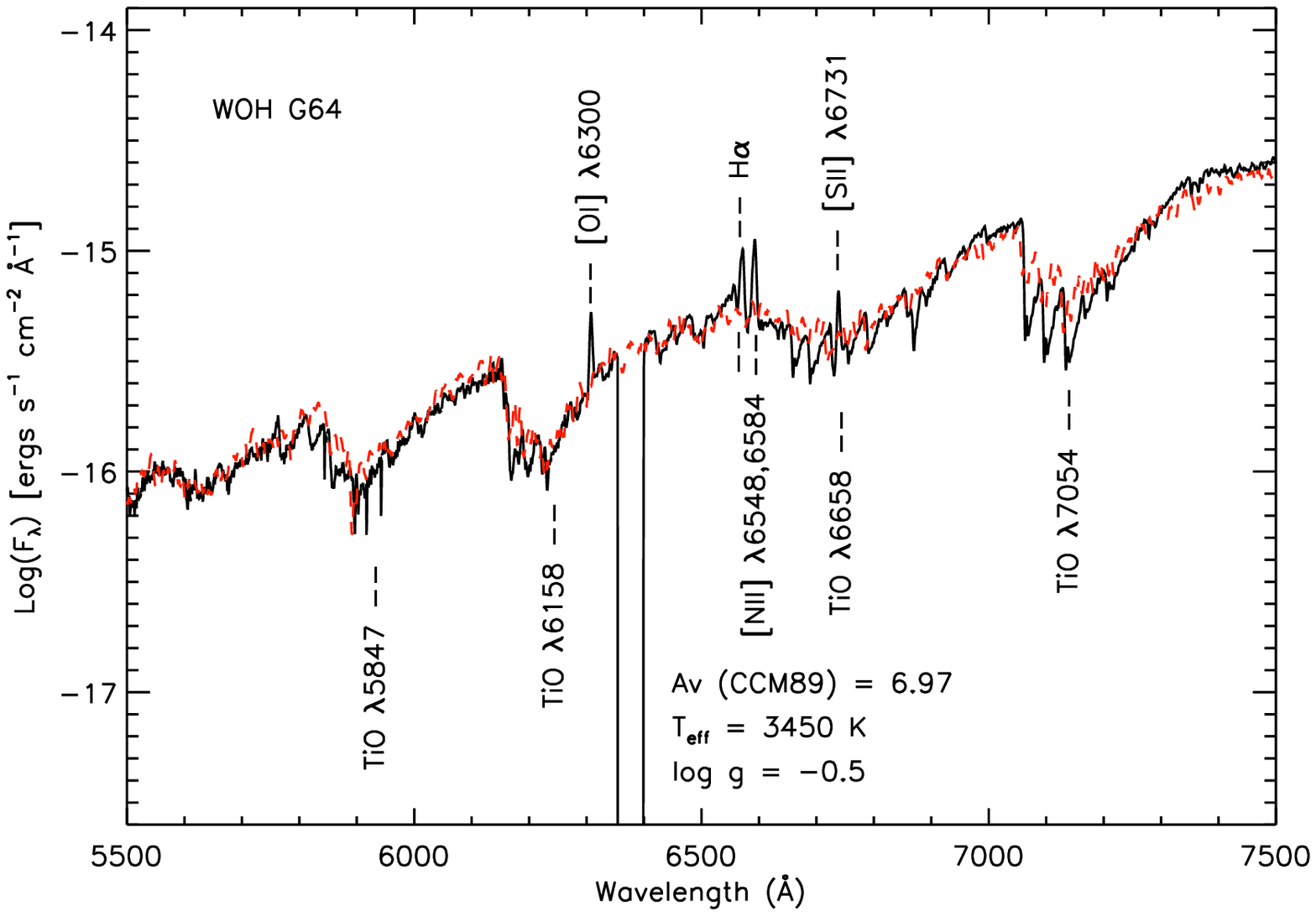}
\plotone{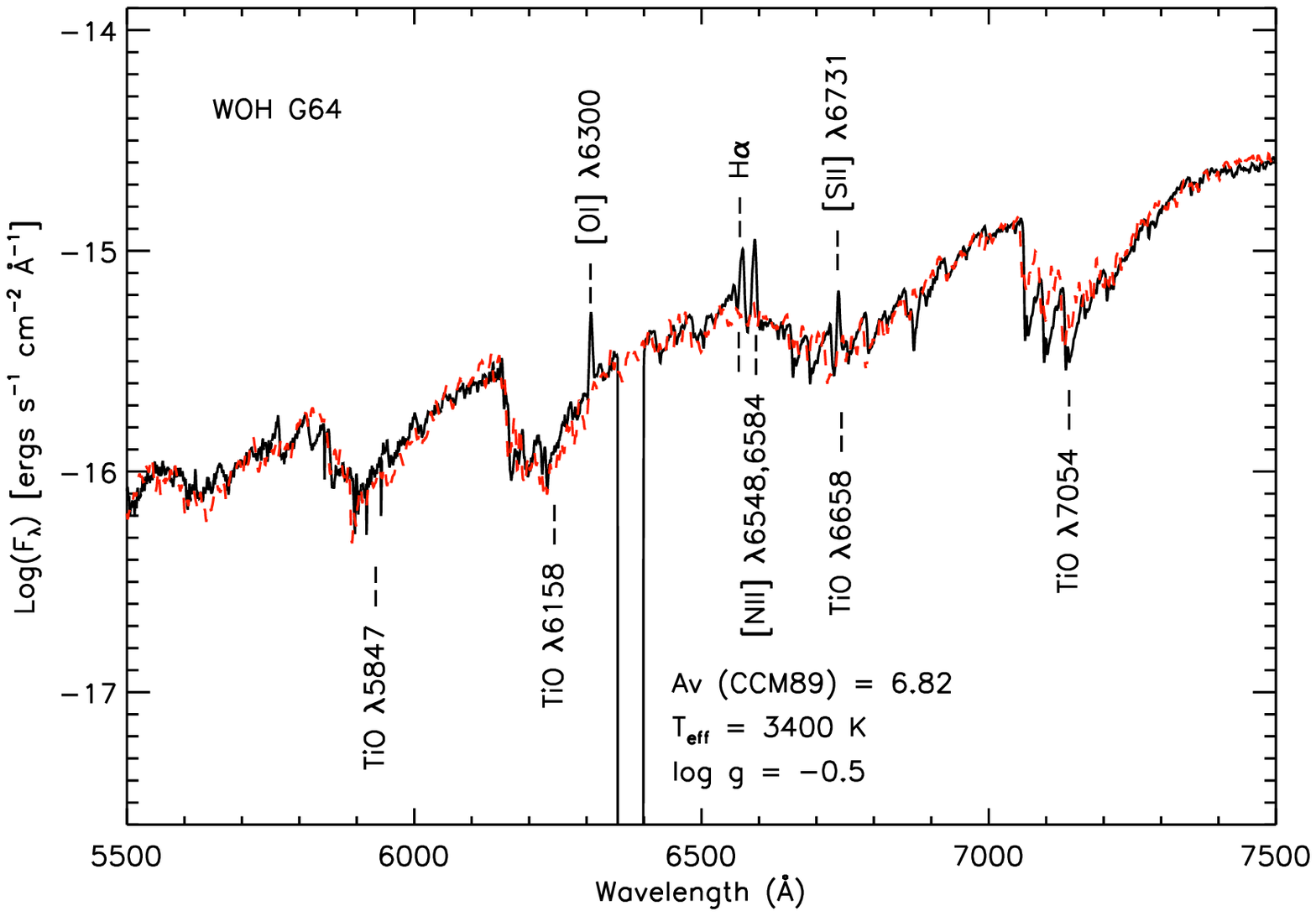}
\plotone{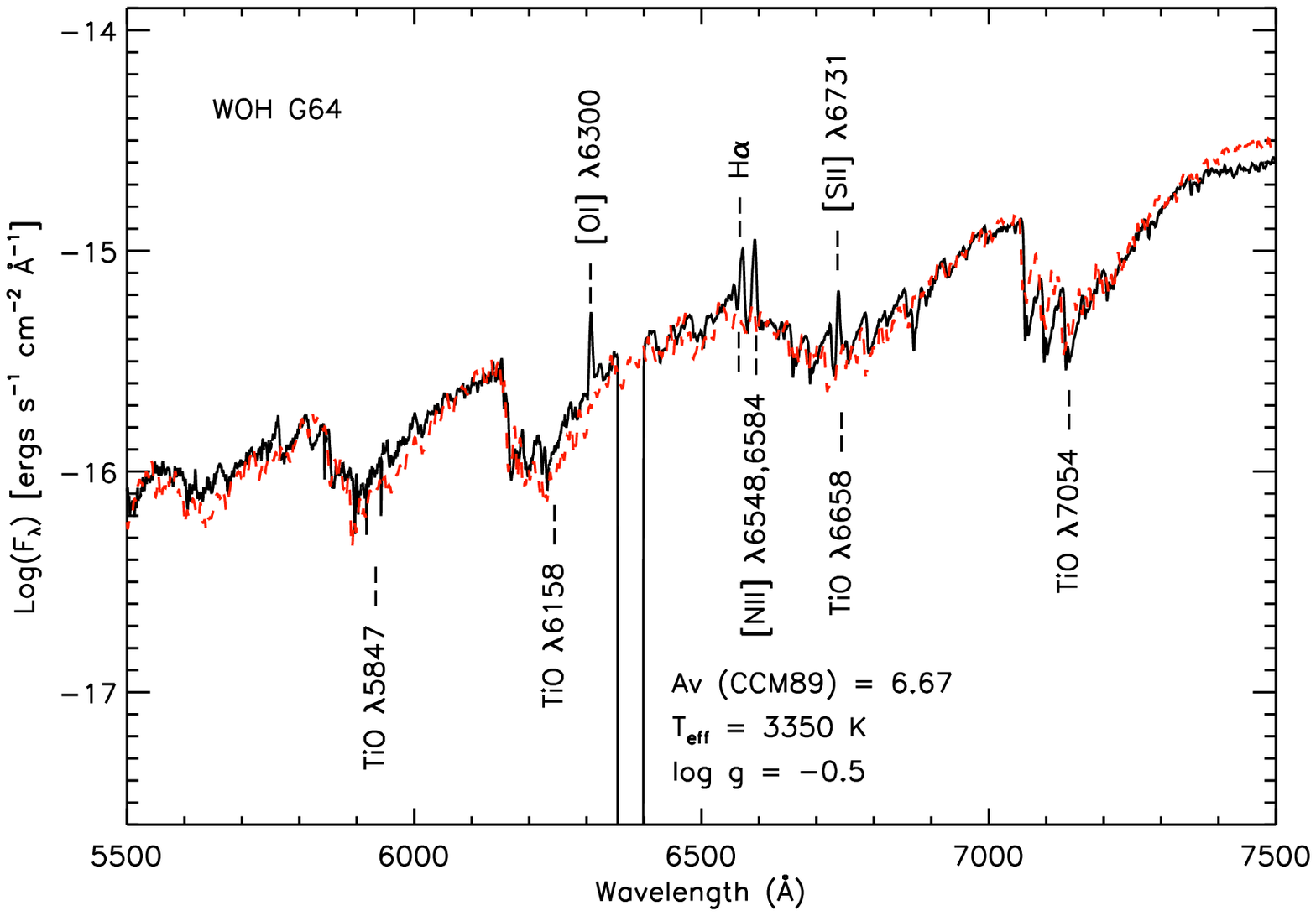}
\caption{\label{fig:spec2} Alternate fits of our spectrum of WOH G64 (black) with the MARCS stellar atmosphere models (green), with the model T$_{\rm eff}$ and $A_V$ increased (top) and decreased (bottom) by 50 K and 0.15 mag, respectively, as compared to our best-fit parameters (center). It is clear in these fits that agreement across all of the TiO bands is poorer for the top and bottom fits, with the 6658\AA\ and 7054\AA\ bands appearing too weak at a T$_{\rm eff}$ of 3450 K, and the 5847\AA, 6158\AA, and 6658\AA\ bands appearing too strong at a T$_{\rm eff}$ of 3350 K. For the center fit, agreement between the observed and model spectrum across all the TiO bands is good. We interpolate this precision to $\pm$ 25 K and take this as the goodness of our fit.}
\end{figure}

From fitting the 5000\AA\ - 9000\AA\ optical spectrum, we find a T$_{\rm eff}$ of 3400 $\pm$ 25 K and $A_V$ = 6.82 $\pm$ 0.15 mag, adopting a log $g$ value of $-0.5$ dex. Our best model fit is shown in Figure \ref{fig:spec}, and alternative model fits demonstrating the precision of these derived properties are shown in Figure \ref{fig:spec2}. In order to confirm that this is the appropriate log $g$ value for WOH G64, we must first determine the star's radius, which in turn requires calculation of the bolometric luminosity. While RSGs in the LMC are known to be variable by up to a magnitude or more in $V$ (Levesque et al.\ 2007), their variability at $K$ is much smaller ($\sim$0.2 mag; Josselin et al.\ 2000). For this reason, we determine the bolometric luminosity based on WOH G64's $K$ magnitude. There are a number of $K$ band observations for this star, including $K_s$ = 6.85 from 2MASS, $K$ = 6.85 from Buchanan et al.\ (2006), $K$ = 6.88 from Elias et al.\ (1986), and $K$ = 6.91 from DENIS (Epchtein et al. 1997). The consistency of these magnitudes confirms that the $K$ magnitude of WOH G64 has stayed quite constant with time. We adopt the 2MASS $K_s$ magnitude for this work, correcting $K_s$ to $K$ by the relation $K = K_s + 0.04$ (Carpenter 2001) and adopting an error of $\pm 0.2$ mag (Josselin et al. 2000). We adopt the T$_{\rm eff}$-dependent bolometric correction at $K$ for the LMC (Paper II):

\begin{equation}
\rm{BC}_K = 5.502 - 0.7392\left(\frac{T_{\rm eff}}{1000 \rm{K}}\right).
\end{equation}

By taking these parameters, calculating $A_K = 0.12 \times A_V$ (Schlegel et al. 1998), and adopting a distance modulus of 18.5 (50 kpc) for the LMC (van den Bergh 2000) with an error of $\pm 0.1$ mag, we determine a physical log $g$ of $-0.7 \pm 0.1$ dex, which agrees with our model surface gravity. With these values, and taking $M_{\rm bol,\odot}$ = 4.74 (Bessell et al.\ 1998) we also determine an M$_{\rm bol} = -9.4 \pm 0.3 $ (log$(L/L_{\odot}) = 5.65 \pm 0.14$), or $\sim$0.6 mag more luminous than the value obtained by Ohnaka et al.\ (2008). We also measure a radius of R/R$_{\odot} = 1970$, with an uncertainty of $\sim$5\%. While van Loon et al.\ (2005) mention some disparity in spectral type determinations between different TiO bands, we find no such disagreement when assigning our spectral type and T$_{\rm eff}$.

\subsection{Properties from $(V-K)_0$}
\label{sec:VK}
As a self-consistency check, we also determine these physical properties based on the $(V-K)_0$ color of WOH G64, as we have done in past papers (Papers I and II, Levesque et al.\ 2007, Massey et al.\ 2009). For a $V$ magnitude we adopt $V$ = 18.63 from the MACHO photometric data (Allsman \& Axelrod 2001). We deredden the photometry using $(V-K)_0 = V - K - (0.88 \times A_V)$ (Schlegel et al. 1998). From this we find that ($V-K$)$_0$ for WOH G64 is 5.74. In Paper II we determined the theoretical relation between T$_{\rm eff}$ and $(V-K)_0$ based on the MARCS models and the assumptions of Bessell et al.\ (1998) concerning the effective broad-band bandpasses, and found

\begin{equation}
T_{\rm eff} = 7621.1 - 1737.74(V-K)_0 + 241.762(V-K)_0^2 - 11.8433(V-K)_0^3.
\end{equation}

With these we determined a T$_{\rm eff}$ of 3372 K. Once again using the bolometric correction at $K$ and $A_K$ and propagating our errors for these values, these numbers in turn produce a log $g = -0.7 \pm 0.1$ dex and M$_{\rm bol} = -9.4 \pm 0.3$ (log$(L/L_{\odot}) = 5.65 \pm 0.13$). Based on these parameters we also calculate a radius of R/R$_{\odot} = 1990$ with an uncertainly of $\sim$5\%. This shows excellent agreement between the parameters derived from spectral fitting as compared to the ($V-K$)$_0$ colors.

The effective temperatures from both spectral fitting and $(V-K)_0$ colors agree with the predictions of the T$_{\rm eff}$ scale for the LMC (Paper II); WOH G64 is the only M5 I RSG known in the LMC, and this additional data point agrees nicely with the general trend that the LMC T$_{\rm eff}$ scale is $\sim$ 50 K cooler than the Galactic scale (the single M5 I RSG in the Milky Way, $\alpha$ Her, has a T$_{\rm eff}$ of 3450 K; Paper I).

The parameters determined from past work, our spectral fitting, and our ($V-K$)$_0$ colors are summarized in Table \ref{tab:params}, where we also compare our values to those derived
by others.

\subsection{The Physical Parameters of WOH G64 and the H-R Diagram}
\label{sec:HR}
In our analysis of WOH G64, we find an unusually high bolometric luminosity,
$M_{\rm bol}= -9.4$ ($\log L/L_\odot=5.65$) compared to the other most luminous
RSGs in the Magellanic Clouds ($M_{\rm bol}=-8.8$).  As shown in Table 1,
our value is consistent
with that of Elias et al.\ (1986) and van Loon et al.\ (2005), but about 0.5~mag more luminous than that found by Ohnaka et al.\ (2008).

As is obvious from the large extinction, WOH G64 is surrounded by a dense
circumstellar envelope.  If this envelope were spherical, then the flux 
absorbed along the line of sight would be re-radiated towards us in the
NIR, but with an asymmetrical circumstelar environment (either a torus or
a disk), we may receive more (or less) than what is absorbed.  If we adopt
the geometrical model proposed by Ohnaka et al.\ (2008), that we view a
torus almost head on, then we are receiving more flux form the combined
system (star plus torus) than is being absorbed.  Based on Figure 5 in Ohnaka et al.\ (2008), we estimate the torus contributes approximately 0.5~mag at $K$.
In order to derive the stellar flux, we then need to correct our
$M_{\rm bol}$ derived from $K$ (-9.4) by this amount, and arrive at
an M$_{\rm bol}=-8.9$.  We list our final adopted parameters   
in Table 1\footnote{Note that Humphreys et al.\ (2007) have argued that
Massey et al.\ (2006) have badly underestimated the actual
luminosity of VY CMa, citing as the ``fundamental astrophysical basis" for their
higher luminosity the integration of the spectral energy distribution, 
most of which is re-radiated thermal emission from the dust around the star.
We note that the the Massey et al.\ (2006) analysis could have been in
error if there were substantial grey absorption, i.e., if the circumstellar dust did not
follow a Cardelli et al.\ (1989) $R_V=3.1$ reddening law, which is certainly
a possibility.  However, it is also true that the argument by Humphreys et al.\ (2007) rests on the unproven (and
unstated) assumption of spherical symmetry for the dust emission.}

We believe our value for the extinction is to be preferred over that
of the Ohnaka et al.\ (2008) analysis, as they adopt a black-body flux
distribution for the star.  The use of a black-body overestimates
the stellar flux in the optical, which leads to an overestimate of the amount of attenuation required to reproduce the
observed data.  In the end, though, our estimates for the bolometric luminosities are in excellent agreement, as shown in Table 1.  Ohnaka et al.\ (2008) did
adopt a somewhat lower T$_{\rm eff}$ than what we found (3200 K vs 3400 K), but
fortuitously they also applied their analysis with a 3400 K T$_{\rm eff}$ and derive
essentially the same bolometric luminosity (Ohnaka 2009, private communication).

With our determination of $T_{\rm eff}$=3400~K and $M_{\rm bol}=-8.9$, we
derive a stellar radius of 1540R$_\odot$.  Given the {\it minimum} formal errors
($\Delta T_{\rm eff}$=25~K and $\Delta M_{\rm bol}=0.1$ mag), the uncertainty of
the radius is 5\%.  This value of $1540 R_\odot$ {\it is} significantly
larger than
the largest Magellanic Cloud stars (1240-1310$R_\odot$), as expected from
its cooler temperature and higher bolometric luminosity.  But, while the
star may be a behemoth by Magellanic Cloud standards, it is about the same
size as the largest RSGs we found in Milky Way (Paper I), where KW Sgr,
Case 75, KY Cyg, and $\mu$ Cep all have essentially the same large radii
as WOH G64.  

\begin{figure}
\plotone{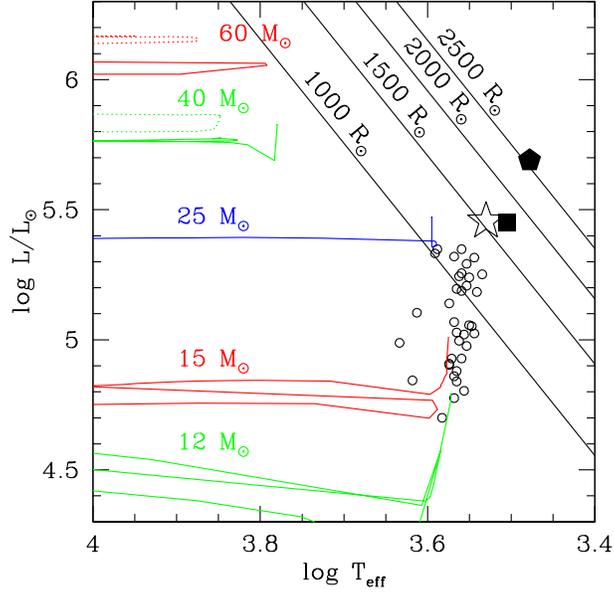}
\caption{\label{fig:hrd} Position of WOH G64 on the HRD, based on the parameters derived in van Loon et al. (2005) (filled pentagon), Ohnaka et al. (2008) (filled square), and the parameters adopted in this paper (open star). The position is compared to the locations of the LMC RSG population presented in Paper II, as well as the evolutionary tracks of the Geneva group. Older, nonrotating evolutionary tracks that include the effects of overshooting are shown as solid lines and come from Schaerer et al. (1993); newer rotating evolutionary tracks, when available, are shown as dotted lines and come from Meynet \& Maeder (2005). Lines of constant radius are shown by diagonal lines in the upper right. It can be seen that our final parameters are in much better agreement with the position of the other LMC RSGs, and assign WOH G64 a notable smaller radius, although WOH G64 remains the coolest, largest, and most luminous RSG in the LMC.}
\end{figure}

In Figure~\ref{fig:hrd} we show the
position of the star in H-R diagram, compared to the location of
the Geneva evolutionary tracks.  We have included too the locations
of the star from van Loon et al.\ (2005) and Ohnaka et al.\ (2008).
We see that the star is cooler than the evolutionary tracks allow,
but that other LMC stars also suffer from this problem; we agree
with the Ohnaka et al.\ (2008) conclusion that this star is on the
edge (or slightly to the right of) of the Hayashi limit.  The luminosity
is in good agreement with the prediction of stellar evolutionary theory,
and we see that the mass corresponding to this luminosity is about $25M_\odot$ (note that although the masses labeled are the stars' {\it initial} masses, mass-loss is expected to remove only a few percent
of a star's mass during its earlier evolution for stars in this mass range.) Combined with
our radius estimate, this leads to a $\log g$ value of -0.5, consistent with the
surface gravity adopted in our fit (\S~\ref{sec:fit}). Given the 5\% uncertainty in
the radius, and adopting a 20\% uncertainty in the mass, leads to an uncertainty of 0.1~dex in
$\log g$.

\section{Emission Lines in the WOH G64 Spectrum}
\label{sec:elines}
Elias et al.\ (1986) report detection of [\ion{O}{1}] $\lambda$6300, H$\alpha$, [\ion{N}{2}] $\lambda$6548, and [\ion{N}{2}] $\lambda$6584 emission lines in their $\sim$6\AA\ resolution spectrum of WOH G64, along with a tentative detection of [\ion{S}{2}] $\lambda\lambda$6717, 6731 emission. Using the [\ion{O}{1}], H$\alpha$, and [\ion{N}{2}] lines they measure a radial velocity of 315 km s$^{-1}$, which implies a redshift compared to the 270-278 km s$^{-1}$ value found in the maser analysis by Marshall et al.\ (2004) and our 294 km s$^{-1}$ measurement from the \ion{Ca}{2} lines (\S~\ref{sec:mem}), given that the Elias et al.\ (1986) value may have a zero-point error as large as 50 km s$^{-1}$. Given that the Elias et al.\ (1986) values have as much as 50 km/sec zero-point error, the significance
of this redshift was not large, but we confirm the redshift below, using more accurate measurements.

\begin{figure}
\epsscale{0.75}
\plotone{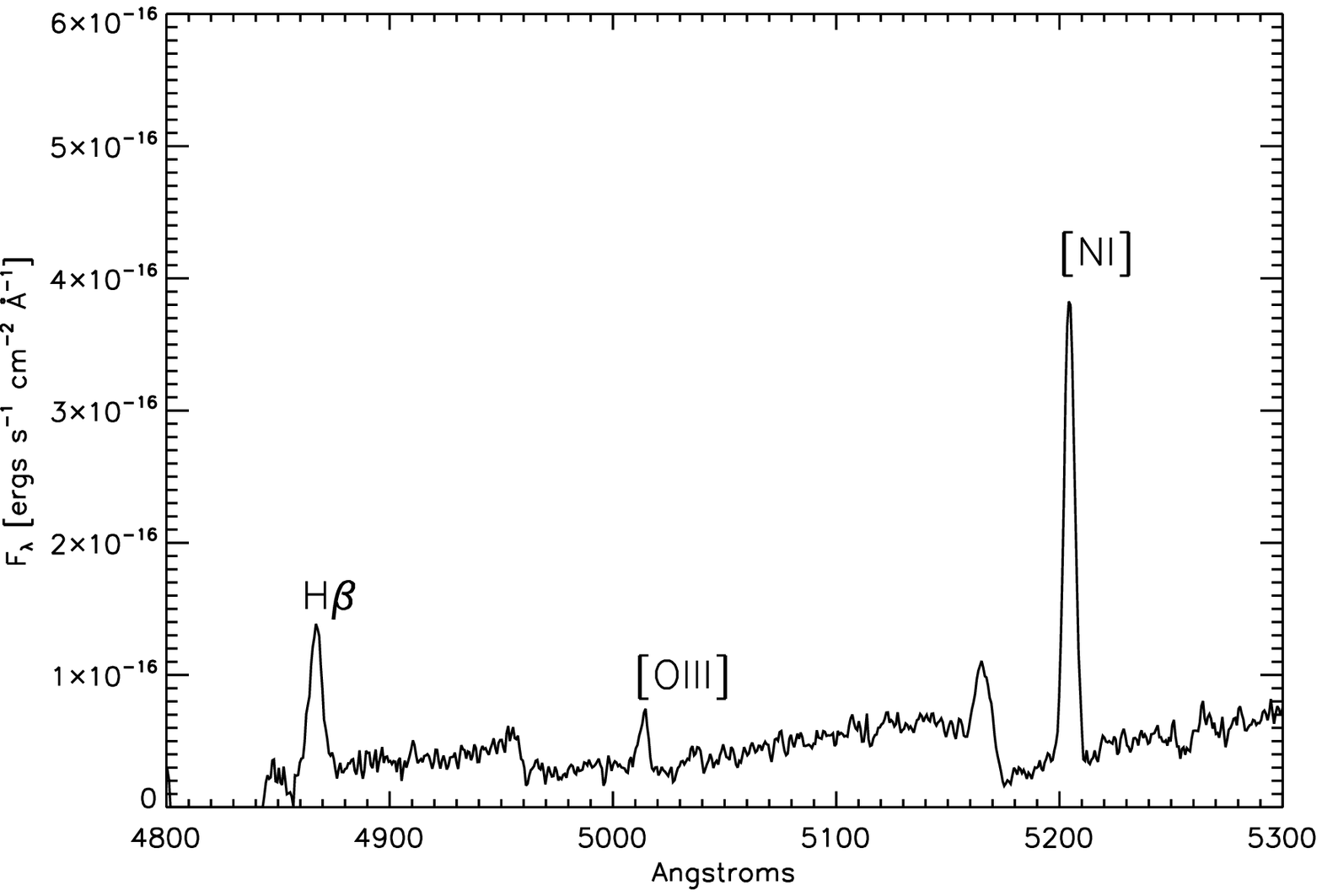}
\plotone{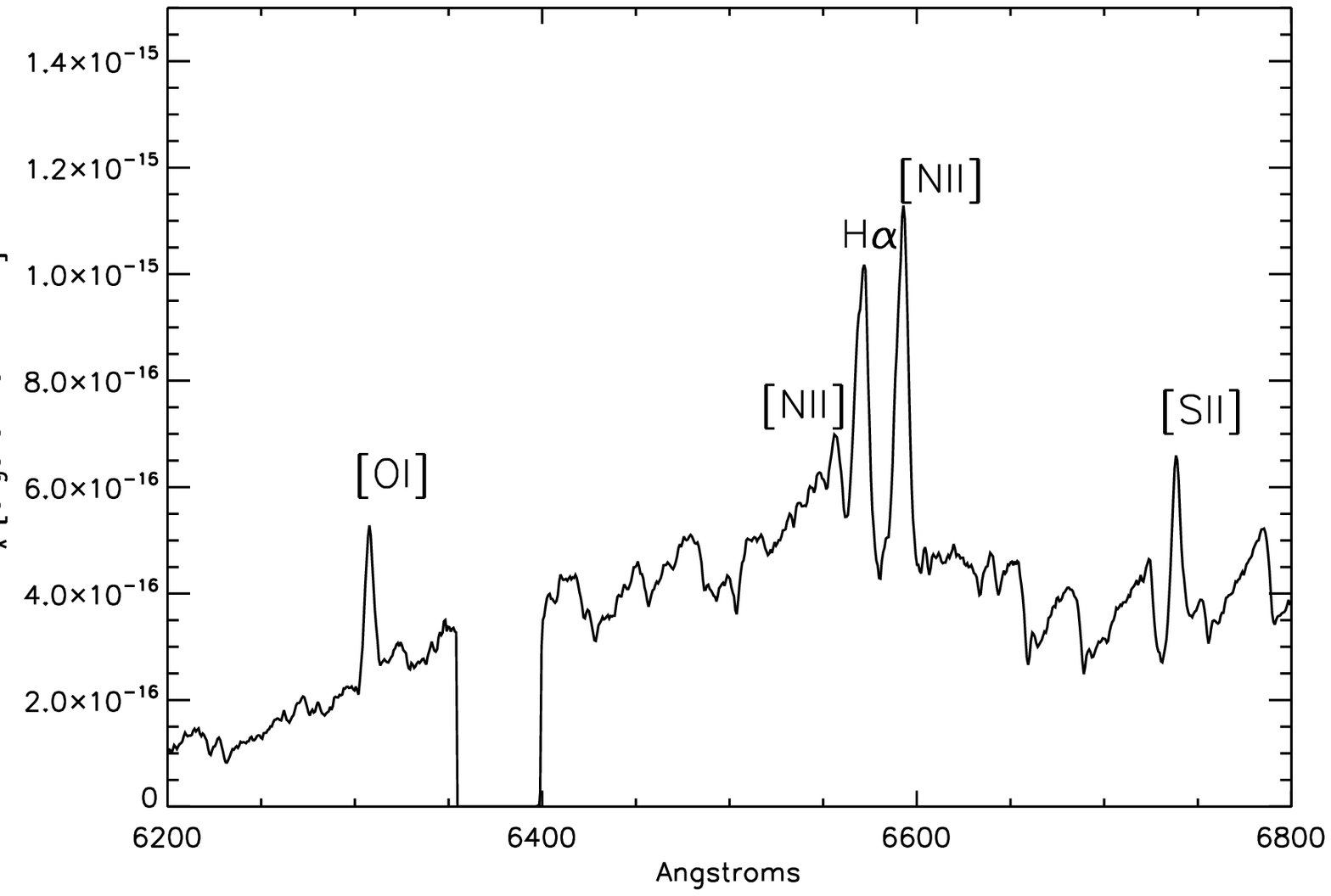}
\caption{\label{fig:elines}Nebular emission line detections in the blue (upper) and red (lower) optical spectrum of WOH G64.}
\end{figure}

In the red part of our spectrum, we detect [\ion{O}{1}] $\lambda$6300, H$\alpha$, [\ion{N}{2}] $\lambda$6548, [\ion{N}{2}] $\lambda$6584, and [\ion{S}{2}] $\lambda$6731 emission (Figure \ref{fig:elines}a).  The [\ion{S}{2}] $\lambda$6717 doublet line is not measurable, as its wavelength is coincident with one of the numerous absorption features in our spectrum. In addition, we detect H$\beta$, [\ion{N}{1}] $\lambda\lambda$5198, 5200, and [\ion{O}{3}] $\lambda$5007 emission in the bluer region of the spectrum (Figure \ref{fig:elines}b). We also note that some of the TiO band heads are in weak emission.  This is also seen in at least one other
RSG with an extreme circumstellar environment, e.g., VY CMa (Hyland et al.\ 1969; Wallerstein \& Gonzalez 2001). The lines are clearly spatially coincident with the stellar spectrum, and there is no evidence of extended (nebular) emission in our good seeing spectra. We would readily detect an extension of 0.3 pixels, corresponding to 0.33", or 0.08 pc at the distance of the LMC. We have measured the radial velocities of the emission lines by fitting Gaussians to them, and find an average (heliocentric) value of $344 \pm 9$ km s$^{-1}$, where the error refers to the standard deviation of the mean; radial velocities for each of these features are given in Table \ref{tab:elines}. This value is significantly larger than the $294 \pm 2$ km s$^{-1}$ we measure for the star itself (\S~\ref{sec:mem}). It is possible that the redshift of these emission lines can be explained in part by scattering of the spectrum by an extended, expanding dust shell, a phenomenon that has previously been observed and described in red supergiants (Romanik \& Leung 1981, and references therein).


We measure the integrated fluxes of these emission lines, and correct these values for extinction based on the H$\alpha$/H$\beta$ emission line ratio, assuming the Balmer decrement for case B recombination (H$\alpha$/H$\beta$ = 2.87, following Osterbrock 1989) and the Cardelli et al. (1989) reddening law. We calculate an $A_V$ = 1.90, very different than the value determined for the star from our spectral fitting; however, it is not surprising that the gas we observe has a net extinction that is lower than the star, and this is in fact consistent with it forming in the outer part of the dust shell. The observed and dereddened integrated fluxes are included in Table \ref{tab:elines}.

With these dereddened fluxes we are able to calculate several common emission line diagnostic ratios, notable log([\ion{N}{2}] $\lambda$6584/H$\alpha$) = 0.05, log([\ion{O}{1}] $\lambda$6300/H$\alpha$) = $-0.42$, and log([\ion{O}{3}] $\lambda$5007/H$\beta$) = $-0.58$), and a lower limit of log([\ion{S}{2}]/H$\alpha$) $= -0.45$ (see Baldwin et al.\ 1981 for a detailed discussion of these diagnostics). The value of the [\ion{O}{1}]/H$\alpha$ ratio is much higher than the minimum value at which [\ion{O}{1}] is considered ``present", defined in Baldwin et al.\ (1981) as log([\ion{O}{1}] $\lambda$6300/H$\alpha$) = $-1.3$. This high relative strength of the [\ion{O}{1}] line implies that the dominant source of ionization is likely shock heating. This is also implied by the relatively high lower limit of the log([\ion{S}{2}]/H$\alpha$) $= -0.45$ ratio in our spectrum (see, for example, Allen et al. 1999, 2008), and we thank the anonymous referee for pointing this out. However, this is not the only explanation, as called to our
attention in thoughtful comments by Nathan Smith (2009, private communication) and
Phil Bennett (2009, private communication).  Bennett argues in particular that
such nebular emission is typical of VV Cep-like systems, i.e., formed by ionization
by a hot companion, but see discussion below.

We also find that nitrogen is highly enhanced in the nebula; if we use the standard galaxy diagnostics of Pettini \& Pagel (2004), which rely upon a constant N/O ratio, we would derive a 12 + log(O/H) = 8.9, considerably in excess of the 12 + log(O/H) = 8.4 that is typical of HII regions in the LMC (Russell \& Dopita 1990).  But, far more likely is the possibility that N is simply enhanced in this gas. Without other line diagnostics (such as [\ion{O}{2}] $\lambda$3727, [\ion{O}{3}] $\lambda$4363, or [\ion{S}{2}] $\lambda$6717) we lack a quantitative knowledge of the oxygen abundance, electron density, and electron temperature. However, such N enrichment is a signature of many shells and rings around massive stars, such as Sher 25 (Brandner et al.\ 1997, Hendry et al.\ 2008). Sher 25 is a blue supergiant, but is thought to be in a blue loop phase, with the high N-enriched gas ejected during a prior evolutionary phase (Brandner et al.\ 1997; see also discussion in Smartt et al.\ 2002 and Hendry et al.\ 2008). Other such examples are also known (Hunter et al. 2008). We note that [\ion{N}{1}] $\lambda \lambda 5198, 5200$ is strongly present.  Like the [\ion{O}{1}] $\lambda 6300$ line, this doublet is also collisionally excited (Osterbrock 1989), and we take its strength as further evidence of shocks and nitrogen enrichment, consistent with the above.

\section{Discussion}
\label{sec:disc}
We have used the MARCS stellar atmospheres and moderate resolution spectrophotometry and broad-band photometry to investigate the physical properties of
the enigmatic supergiant WOH G64.  In our addition our study conclusively shows that the star is
a supergiant member of the LMC, rather than (say) an unusual giant star in our own
galaxy's halo.

Our analysis finds an effective temperature of 3400 $\pm$ 25 K, considerably warmer than
previous studies have adopted. With our parameters from spectral fitting, and taking into account the additional flux contribution at $K$ from the asymmetric circumstellar dust envelope as seen in Ohnaka et al. (2008), we determine M$_{\rm bol} = -8.9$ for WOH G64. While WOH G64 has previously been cited as having a T$_{\rm eff}$ and M$_{\rm bol}$ that would make it the largest RSG known, those parameters have also placed it at odds with the position of the evolutionary tracks on the H-R diagram. By contrast, our parameters place WOH G64 closer to the predictions of stellar evolutionary theory and in good agreement with other RSGs in the LMC. We determine a stellar radius of $R/R_{\odot}$ = 1540 which, while still the largest in the LMC, is now in good agreement with the size of other large RSGs ($\sim1420 - 1520$R$_{\odot}$; Paper I), given that such radii have 5\% errors or larger.

The question still remains as to whether or not WOH G64 is a Levesque-Massey variable. Its extreme physical properties and position on the H-R diagram are similar to those observed in Levesque-Massey variables. However, there is currently no evidence for spectral variability based on past observations; spectra obtained by Elias et al.\ (1986), van Loon et al.\ (2005), and this work all show WOH G64 remaining in a consistent, cool, late-M-type state. The disagreement in type between M5 and M7-8 is almost certainly a result of differing methods rather than changes in the spectrum. However, more frequent observations of WOH G64 could well uncover variations in its $V$ magnitude and T$_{\rm eff}$; data from the All Sky Automated Survey (ASAS) project (Pojmanski 2002) demonstrates that the star's $I$ magnitude shows considerable variability, but there is no $V$ or $K$ band data available. On the other hand, the star's very high reddening, well-defined dusty torus, and considerable maser activity would be unique among the Levesque-Massey variables; these properties suggest WOH G64's similarity to the Milky Way's VY CMa, as well as other dust-enshrouded RSGs with maser activity such as NML Cyg, VX Sgr, and S Per (Schuster et al.\ 2006).

Finally, we detect a number of emission lines in the WOH G64 optical spectrum, including H$\alpha$ and H$\beta$, [\ion{O}{1}], [\ion{N}{1}], [\ion{S}{2}], [\ion{N}{2}], and [\ion{O}{3}]. The radial velocity of these features, 344 $\pm$ 9 km s$^{-1}$, is considerably greater than the star's radial velocity, 294 $\pm$ 2 km s$^{-1}$, which may be due to scattering by the expanding dust region. We also note the presence of some TiO band heads in emission. This has been previously seen in the spectrum of VY CMa, which is due to the extreme circumstellar environment. Inspection of our data shows that the emission is not spatially extended, and the strength of the [\ion{O}{1}] feature suggests that the dominant source of ionization for these features is shock heating.

We cannot reject the possibility that WOH G64 is a binary system, with the nebular component contributed by a hot companion, but we do not favor this interpretation. First is the issue of the nebular radial velocities.  We find a significant (50 km s$^{-1}$) velocity difference between the nebular emission and the stellar component, and, at first blush, this might be taken as an indication of binarity; i.e., with the nebula emission (somehow) tracking the motion of the hot component. However, a similar velocity was also noted by Elias et al.\ (1986).  While this could be a coincidence, it certainly doesn't argue in favor of a binary hyptothesis. Besides, in a short-period system we would not expect the ionized gas to be ``attached" to the hot star. Rather, we prefer the explanation that the nebular redshift is the result of scattering within the expanding dust shell.  We do note that if the putative hot companion shared the extinction measured from the nebular lines, rather than that of the RSG, its flux would dominate even in the red part of the spectrum, which is clearly not the case. If instead the companion shared the same high extinction as the RSG, but its HII region did not, we would still expect that the companion would dominate the flux in the blue part of the spectrum. A better spectrum going further into the blue should be able to substantiate
or refute the possibility that a hot companion is present with the same extinction, as a late O-type dwarf 
would have a bolometric luminosity of $\log L/L_\odot \sim 5$ ($M_{\rm bol}\sim -7.5$).  Accounting for the bolometric corrections involved would lead to $M_V=-4$ for the companion, and $-6$ for the RSG.  Thus, at $B$, the two stars would be equally bright, as $(B-V)_0~\sim -0.3$ for a late-type O star, while
$B-V\sim 2$ for a RSG (see Massey 1998a, 1998b, and Paper I). However, further investigation into the source of these emission lines will likely require further spectral analyses, as well as observations of WOH G64 in the bluer regions of the spectrum.

\acknowledgements
We gratefully acknowledge LCO's hospitality and assistance provided during our observations for this paper. This paper made use of data from the Two Micron All Sky Survey (2MASS), which is a joint project of the University of Massachusetts and the Infrared Processes and Analysis Center, California Institute of Technology, funded by the National Aeronautics and Space Administration and the National Science Foundation. We thank Phil Bennett, George Herbig, Deirdre Hunter, Eric Josselin, Scott Kenyon John Monnier, Keiichi Ohnaka, Nathan Smith, Jacco van Loon, George Wallerstein, and the anonymous referee for helpful comments and discussion regarding this manuscript. E. L.'s participation was made possible in part by a Ford Foundation Predoctoral Fellowship. This work was supported by the National Science Foundation through AST-0604569 to P. M.

\begin{deluxetable}{l c c c c c c}
\rotate
\tablewidth{0pc}
\tablenum{1}
\tablecolumns{7}
\tablecaption{\label{tab:params} Physical Parameters of WOH G64}
\tablehead{
\colhead{Reference}
& \colhead{Spectral Type}
& \colhead{T$_{\rm eff}$ (K)}
& \colhead{M$_{\rm bol}$ $\left({\rm log} \frac{L}{L_{\odot}}\right)$}
& \colhead{$\frac{R}{R_{\odot}}$}
& \colhead{$A_V$}
& \colhead{log $g$}
}
\startdata
Elias et al.\ 1986 & M7.5 &\nodata &-9.6 (5.75)&\nodata &$\sim$3 &\nodata \\
van Loon et al.\ 2005 & M5, M7-8\tablenotemark{a} &3008 &-9.5 (5.70)&2575\tablenotemark{b} &\nodata &\nodata \\
Ohnaka et al.\ 2008 & \nodata &3200\tablenotemark{c} &-8.9 (5.45)&1730 &$\sim$10\tablenotemark{d} &\nodata \\
Ohnaka et al.\ 2008 & \nodata &3400\tablenotemark{c} &-8.9 (5.45)\tablenotemark{e}&1730\tablenotemark{e} &?\tablenotemark{e} &\nodata \\
This work, ($V-K$)$_0$ colors& M5 &3372 &-9.4 (5.65)&1990 &6.82 &-0.7  \\
This work, spectral fitting & M5 &3400 &-9.4 (5.65)&1970 &6.82 &-0.7 \\
This work, final preferred parameters & M5 &3400 &-8.9 (5.45)&1540 &6.82 &-0.5 \\
\enddata
\tablenotetext{a} {Different spectral types determined based on different measured line depths.}
\tablenotetext{b} {Calculated based on listed M$_{\rm bol}$ and T$_{\rm eff}$.}
\tablenotetext{c} {Adopted based on previously-assigned spectral types.}
\tablenotetext{d} {Formally 9.8$\pm$2.2, but the high value comes from assuming a blackbody distribution for the star with a low T$_{\rm eff}$; see text.}
\tablenotetext{e} {Adopted based on reportedly good agreement with parameters derived from T$_{\rm eff}$ = 3200 K; Ohnaka et al. (2008)}
\end{deluxetable}

\clearpage

\begin{deluxetable}{l c c c}
\tablewidth{0pc}
\tablenum{2}
\tablecolumns{4}
\tablecaption{\label{tab:elines} Emission Line Radial Velocities (RVs) and Fluxes}
\tablehead{
\colhead{Line ID}
& \colhead{$RV$}
& \colhead{Observed Flux}
& \colhead{Corrected Flux} \\
\colhead{}
&\colhead{(km s$^{-1}$)}
&\colhead{(ergs cm$^{-2}$ s$^{-1}$) $\times$ 10$^{-15}$}
&\colhead{(ergs cm$^{-2}$ s$^{-1}$) $\times$ 10$^{-15}$}
}
\startdata
H$\beta$ & 330 & 0.82 & 6.5\\
$\rm {[OIII]}$ 5007 & 306 & 0.24 & 1.7\\
$\rm {[NI]}$ 5199/5202\tablenotemark{a} &\nodata &2.12 &13.9 \\
$\rm {[OI]}$ 6300 & 336 &1.53 &6.7\\
$\rm {[NII]}$ 6548 & 371 &0.72 &3.0\\
H$\alpha$ & 355 &4.31 &17.8\\
$\rm {[NII]}$ 6584 & 385 &4.86 &20.0 \\
$\rm {[SII]}$ 6731 & 326 &1.59 & 6.3\\
\enddata
\tablenotetext{a} {Blend; no radial velocity calculated.}

\end{deluxetable}

\end{document}